\documentclass[prl,aps,preprint,superscriptaddress]{revtex4}
\usepackage{graphicx}
\usepackage{epsfig}
\usepackage{amssymb}
\usepackage{amsmath}

\def\ba{\begin{eqnarray}}
\def\ea{\end{eqnarray}}
\def\bse{\begin{subequations}\begin{align}}
\def\ese{\end{align}\end{subequations}}
\def\sweet{\mc{E}_{DC}^{\textrm{off, sweet}}}

\newcommand{\mc}[1]{\mathcal{#1}}

\newcommand{\bra}[1]{\langle#1|}
\newcommand{\ket}[1]{|#1\rangle}

\begin{document}
\title{Polar molecules near superconducting resonators: a coherent, all-electrical, molecule-mesoscopic interface}

\author{A. Andr\'{e}}
\affiliation{Department of Applied Physics, Yale University, New Haven, Connecticut 06520, USA}
\affiliation{Department of Physics, Harvard University, Cambridge, Massachusetts 02138, USA}
\author{D. DeMille}
\affiliation{Department of Physics, Yale University, New Haven, Connecticut 06520, USA}
\author{J. M. Doyle}
\affiliation{Department of Physics, Harvard University, Cambridge, Massachusetts 02138, USA}
\author{M. D. Lukin}
\affiliation{Department of Physics, Harvard University, Cambridge, Massachusetts 02138, USA}
\author{S. E. Maxwell}
\affiliation{Department of Physics, Harvard University, Cambridge, Massachusetts 02138, USA}
\author{P. Rabl}
\affiliation{Institute for Theoretical Physics, University of Innsbruck}
\author{R. J. Schoelkopf}
\affiliation{Department of Applied Physics, Yale University, New Haven, Connecticut 06520, USA}
\affiliation{Department of Physics, Yale University, New Haven, Connecticut 06520, USA}
\author{P. Zoller}
\affiliation{Institute for Theoretical Physics, University of Innsbruck}
\affiliation{ Institute for Quantum Optics and Quantum Information of the Austrian Academy of Sciences, A-6020 Innsbruck, Austria}

\begin{abstract}

The challenge of building a scalable quantum processor requires
consolidation of the conflicting requirements of achieving coherent
control and preservation of quantum coherence in a large scale
quantum system. Moreover, the system should be compatible with miniaturization
and integration of quantum circuits. Mesoscopic solid state systems such as
superconducting islands and quantum dots feature robust control
techniques using local electrical signals and self-evident scaling
based on advances in fabrication; however, in general the quantum states of solid state devices tend to decohere rapidly. In contrast, quantum optical systems based on trapped ions and neutral atoms exhibit
dramatically better coherence properties, while miniaturization of
atomic and molecular systems, and their integration with mesoscopic
electrical circuits, remains an important challenge. Below we
describe methods for the integration of a single particle system -- an isolated polar molecule -- with
mesoscopic solid state devices in a way that produces robust,
coherent, quantum-level control. The methods described include the
trapping, cooling, detection, coherent manipulation and quantum coupling of
isolated polar molecules at sub-micron dimensions near cryogenic
stripline microwave resonators.  We show that electrostatically trapped polar molecules can exhibit strong confinement and fast, purely electrical gate control.  Furthermore, the effect of electrical noise sources, a key issue in quantum information processing, can be suppressed to very low
levels via appropriate preparation and manipulation of the polar
molecules. Our setup provides a scalable cavity QED-type quantum
computer architecture, where entanglement of distant qubits stored
in long-lived rotational molecular states is achieved via exchange of
microwave photons.

\end{abstract}
\maketitle


Polar molecules \cite{doyle2004} have exceptional features for use in quantum information systems. Stable internal states of polar molecules can be controlled by electrostatic fields, in analogy with quantum dots and superconducting islands \cite{loss1998,wallraff:060501,makhlin2001,nakamura1999}. 
This controllability of polar molecules is due to their rotational degree of freedom in combination with the asymmetry of their structure (absent in atoms). By applying moderate laboratory electric fields, rotational states with transition frequencies in the microwave range are mixed, and the molecules acquire large dipole moments (on the order of a few Debye, similar to the transition dipole moments of optical atomic transitions).  These dipole moments are the key property that makes polar molecules effective qubits in a quantum processing system. Furthermore, application of electric field gradients leads to large mechanical forces, enabling trapping of the molecules. Finally, polar molecules also combine the large-scale preparation available for neutral atoms \cite{folman2002,mandel2003} with the ability to move them with electric fields, as with ions 
\cite{ciracphystoday2004,leibfried2005,haffner2005}. 

We here show that trapping the molecules at short distances from a superconducting transmission line resonator greatly enhances the coupling of the molecular rotational transitions to microwave radiation, leading to methods both for cooling the molecule and for manipulation of the molecule as a qubit. The qubit can be encoded in rotational states and coherently transferred to the resonator. It has already been shown that the strong free-space dipole-dipole coupling of polar molecules is viable for construction of a quantum computer\cite{demille:067901}. Here we show that coupling to and through a resonator enables high fidelity quantum gates at a distance as well as cooling, high fidelity readout, and the construction of a potentially scalable quantum information processor.


{\it Chip-based microtraps for polar molecule qubits}

In a polar molecule, the body-fixed electric dipole moment (\mathversion{bold}$\mu$\mathversion{normal}) gives rise to large transition moments between rotational states, which are separated by energies corresponding to frequencies in the microwave range.  
The level structure of a diatomic rigid rotor is determined by the Hamiltonian $H_{rot} = \hbar B\bold{N}^2$, where $\bold{N}$ is the rotational angular momentum and B is the rotational constant.  This Hamiltonian gives energy levels $E_N = \hbar BN(N+1)$ that are $(2N+1)$-fold degenerate, corresponding to the different projections $m_N$.  In the presence of a DC electric field \mathversion{bold}$\mc{E}$\mathversion{normal}$_{DC} = \mc{E}_{DC}\hat{z}$, the Stark Hamiltonian $H_{St} = -$\mathversion{bold}$\mu$\mathversion{normal}$\cdot$\mathversion{bold}$\mc{E}$\mathversion{normal}$_{DC}$ mixes rotational states and induces level shifts that break the degeneracy between states of different $|m_N|$.  Two of the low-lying states, $\ket {1}\equiv \ket{N=1} \equiv \ket{N=1, m_N=0}$ and $\ket{2} \equiv \ket{N=2} \equiv \ket{N=2, m_N=0} $ with corresponding eigenenergies $E_1$ and $E_2$, are weak-field seeking; i.e., their energy increases with larger $\mc{E}_{DC}$.  We take these two states as the working rotational qubit states for the system. As discussed below, qubits stored in spin or hyperfine
states can be transferred to these rotational qubits with microwave
pulses. A static electric field minimum can be formed in free space, allowing for trapping of these low-field-seeking molecules.

The maximum trap depth possible, determined by the maximum Stark shift of the $\left| 1 \right\rangle$ state, is $U_{max} \approx 0.6 \hbar B$, attained at an electric field $\mc{E}_{DC}^{max} \sim 6 \hbar B/\mu$.  The energy splitting between rotational states, $\omega_0$, is field dependent due to mixing with other $\ket{N, m_N}$ states. However, under all conditions discussed in this paper the relation $\omega_0 \equiv (E_2-E_1)/\hbar =  4B$ is valid to within 10\%. 
Explicit calculations of the molecular energy levels as a function of $\mc{E}_{DC}$ are shown in Fig. 1.  For concreteness, throughout the paper we use CaBr as our example molecule, although many others share the desired properties for quantum computation in our scheme.  For CaBr, $\mu = 2\pi \times 2.25$ MHz V$^{-1}$ cm (4.36 Debye), $B = 2\pi \times 2.83$ GHz, $U_{max} \approx 80$ mK, $\mc{E}_{DC}^{max} \approx 7 \times 10^3$ V cm$^{-1}$ and $\omega_0 =2 \pi \times 11.3$ GHz. 

A variety of macroscopic electrostatic traps for polar molecules have been proposed and/or implemented \cite{bethlem2000,rieger:173002,xia2005}.  We here describe a novel Electrostatic Z-trap (EZ trap, Fig.~\ref{fig:2}a), a mesoscopic electric trap that is closely related to the magnetic Z trap\cite{folman2002}, which is widely used in miniature atomic traps.  The EZ trap creates a non-zero electric field minimum in close proximity to the surface of a chip. The field at the bottom of the trap is designated $\mc{E}_{DC}^{\textrm{off}}$. Metallic electrodes set to the appropriate DC voltages give rise to the inhomogeneous trapping field.  Adjustment of the trap and bias electrodes sets the trap depth as well as $\mc{E}_{DC}^{\textrm{off}}$ and the position of the trap center, typically at height $z_0 \sim w$ above the surface, where $w$ is the typical spacing between the trap electrodes.  

With EZ trap electrodes thin compared to $w$ and held at voltage $V_{EZ}$, maximum DC electric fields $\mc{E}_{DC}^{max} \sim V_{EZ}/w$ can be generated,  leading to harmonic trap potentials with depth $U_0 \sim 0.1~ \mu \mc{E}_{DC}^{max}$ and motional frequency $\omega_t \sim \sqrt{2U_0/(mw^2)}$ for a molecule of mass $m$.  With electrode dimensions ranging from $w = 0.1 - 1~\mu$m and corresponding voltages $V_{EZ} \sim 0.1-1$ V, $U_0 \sim U_{max}$ and $\omega_t \sim 2 \pi \times 6 - 0.6$ MHz for CaBr.  Note that the trapping potential is only slightly modified by the van der Waals interactions of the molecules with the chip surface (see Methods).

To efficiently load the EZ trap, it is necessary to have a source of cold polar molecules that can achieve phase-space density $\Phi$ comparable to that for a single molecule in the EZ trap, i.e., $\Phi \sim w^{-3}U_0^{-3/2} (\sim 10^{15}$ cm$^{-3}$K$^{-3/2}$ for our nominal conditions).  While this value is beyond what has yet been achieved with polar molecules, it seems feasible with several methods currently under development. For example, pre-cooled molecules (produced using, e.g., techniques such as buffer-gas cooling \cite{egorov:043401} or Stark slowing \cite{bethlem2000}) could be trapped and then further cooled to increase $\Phi$, using a variety of techniques. Possibilities include evaporative\cite{weinstein2002, demille04, masuhara88} or sympathetic cooling, cavity cooling\cite{vuletic:033405}, or the stripline-assisted side-band cooling described below. With sufficient phase space density, mode matching to a micron-sized EZ trap configuration should be possible using electrostatic fields \cite {bochinski2003}.

{\it Cavity QED with polar molecules and superconducting striplines}

Superconducting stripline resonators \cite{wallraff2004} can be used to confine microwave fields to an extremely small volume \cite{blais:062320}, $V \sim w^2 \lambda\ll \lambda^3$, where $\lambda$ is resonant wavelength. One important consequence is the large vacuum Rabi frequency $g=\wp \mc{E}_{0} /\hbar$ for molecules located close to such a resonator, enabling coherent coupling of the molecule to the quantum state of the resonator field.  (Here $\wp$ is the transition dipole matrix element and $\mc{E}_{0} \propto V^{-1/2}$ is the zero-point electric field; $\wp \approx 0.5\mu$ under the relevant conditions.)

When the microwave field is confined to a resonator and is quantized, the coupling becomes the well-known \cite{scullyQO} Jaynes-Cummings Hamiltonian $\hat{H}=-\hbar g(\hat{a}^\dagger\hat{\sigma}^-+\hat{a}\hat{\sigma}^+)$, where $\hat{a}$ is the annihilation operator for the resonator mode, 
$\hat{\sigma}^- = \ket{1}\bra{2}$ is the lowering operator for the molecule, and $\ket{1},\ket{2}$ are the two states coupled by the field.  The value of $g$ can be 
calculated as follows (see also \cite{sorensen:063601}). 
The stripline zero-point voltage is $V_{0} = \sqrt{\hbar\omega /(2C)}$, where $C$ is the effective capacitance of the stripline resonator.  For impedance matching with standard microwave devices, $C = \pi/(2 \omega Z_0)$, with $Z_0 = 50 \Omega$.  At a height $z_0$ above a stripline with conductor spacing $w$, the zero-point electric field is $\mc{E}_{0} \sim f(z) V_{0}/w$.  Here $f(z)$ is a dimensionless geometric factor describing the reduction of the field away from the electrodes; $f(z) = 1$ for $z \ll w$ and we find through simulation that $f(z) \sim 0.5 (w/z)$ for $z < w$.
For CaBr trapped at a height $z \lesssim w = 0.1-1~\mu$m above the resonator, the single photon Rabi frequency in the range  $g/2\pi\simeq 400 - 40$ kHz.

High $Q$ resonators (internal $Q$'s of $10^6$ have been demonstrated
 \cite{wallace:1754,wallraff2004,frunzio}) lead to long microwave photon lifetime.  When coupling to the resonator is stronger than the cavity decay rate (
$\kappa=\omega/Q$), coherent quantum state exchange between the polar molecule and the resonator field is possible.  That is, for molecules held close to a small resonator, the electric dipole interaction with the resonator mode is strong enough to reach the strong coupling regime of cavity QED \cite{raimond2005,miller2005}.

{\it Cooling via resonator-enhanced spontaneous emission}

Strong coupling of the molecule to the microwave cavity enhances spontaneous emission of excited rotational states, which can be employed to cool the trapped molecules.
When molecules are initially loaded into the trap, their temperature may be as high as the trap depth $U_0$, corresponding to a mean number of trap excitations $\bar{n}_{trap} \sim U_0/(\hbar \omega_t)$. Depending on the cold molecule production method employed, $\bar{n}_{trap}$ can be very large. As discussed below, the best coherence properties for molecular superposition states are achieved when the molecule is in the ground motional state of the trap. Hence, it is desirable to cool the motional degree of freedom.
The tight confinement of the molecule in the EZ trap suggests side-band cooling, as done with trapped ions \cite{wineland79} . If the molecule were in free space this would not be possible due to the low natural decay rate of excited rotational states, $\gamma < 10^{-5}$s$^{-1}$. 
However, the cavity-enhanced spontaneous emission makes sideband cooling feasible.

The absorption spectrum of the trapped molecule in the combined rotational ($N=1,2$) and motional ($n$) states , $\ket{N, n}$, consists of a carrier at frequency $\omega_0$ and sidebands at frequencies $\omega_0+(m-n)\omega_t$ ($n,m=0,1,\cdots$). Electromagnetic coupling to sidebands arises due to the position dependence
of $g$.  Sideband cooling occurs when a driving field is tuned to the lower energy sideband $\ket{1,n}\rightarrow\ket{2,n-1}$, while the resonator mode is resonant with the $\ket{2,n}\rightarrow\ket{1,n}$ transition (Fig.~\ref{fig:3}a). Each scattering cycle reduces the kinetic energy of the molecule by $\hbar\omega_t$. 
If $\kappa < \omega_t$, the sidebands can be resolved.    
The lower energy sideband state can either be directly excited by an on-resonant driving field or virtually excited by a red-detuned field. De-excitation occurs by emitting a photon into the resonator, which can then decay out of the resonator (see Methods).
The maximum cooling rate can be estimated as follows.  For $g \ll \kappa~(g \gtrsim \kappa)$, the cavity-enhanced spontaneous emission rate for the cavity tuned to resonance, $\omega = \omega_0$, is given by $\Gamma_c = g^2/\kappa~(\Gamma_c = \kappa/2)$.  For example, with $g=2\pi\times 40-400$ kHz and $\kappa=2\pi\times 10$ kHz ($Q=10^6$), $\Gamma_c \sim 2\pi\times 5$ kHz;  this yields a cooling rate $dn/dt = \Gamma_c$ and hence $dE/dt = \hbar \omega_t \Gamma_c~ \approx 10$ K/s for trapping frequency $\omega_t \sim 2\pi \times 5 $ MHz. Thus, a trapped molecule can be cooled to the motional ground state of the trap with a rate much higher than the observed trap loss rates of atoms from microtraps -- typically limited by background gas losses.

In the absence of any heating mechanisms, cooling proceeds until the molecule's mean motional quantum number in the trap, $\bar{n}_t$ equals the mean number of thermal microwave photons in the resonator mode, $\bar{n}_{\gamma}$. Due to the frequency mismatch between the resonator mode frequency $\omega$ and the trap frequency $\omega_t$, the effective final temperature $T_t$ of the trap degree of freedom is lower than the ambient resonator temperature, $T_r$, by the large factor  $R = \frac{\omega_t}{\omega}$. For example, $T_t  < 100 ~\mu$K for $T_r = 100$ mK.  
Technical noise sources may lead to a larger photon occupancy in the cavity than the nominal thermal value $\bar{n} = \textrm{exp}(-\hbar \omega/k T_r) \sim 5 \times 10^{-3}$. However, it has been shown that $\bar{n}_{\gamma} \ll 1$ is achievable\cite{wallraff2004}.

The location of the EZ trap center is determined by the voltage on the electrodes (Fig.~\ref{fig:2}a), so that fluctuations of this voltage cause random motion of the trap center and heating of the molecule's motion in the trap.  As a worst case, we can assume that the micron-sized electrodes experience the typical $1/f$ charge noise as measured in work with single-electron transistors and charge qubits. These may be roughly analogous (at micron size and mK temperatures) to the ''patch potentials'' observed in ion traps\cite{deslauriers:043408}. The typical charge noise\cite{Zorin96}
density $S_Q$ has a $1/f$ dependence, with magnitude $S_Q(f)=10^{-6}- 10^{-7} e^2/f$; on a metal electrode with $C_t\sim 1$ fF (typical for $\mu$m-scale features, the corresponding voltage spectral density is $S_V(f) = S_Q/C_t^2 \sim \frac{10^{-14}}{f}\ {\rm V}^2{\rm Hz}^{-1}$.
With the trap operating in the linear Stark regime, the heating rate, defined as the rate of excitation from $\ket{0}$ to $\ket{1}$, is $\Gamma_{01} \sim \omega_t^2 (w/a_0)^2 \frac{S_V(\omega_t)}{V_{EZ}^2}$ Hz, where $a_0=\sqrt{\hbar/(2m\omega_t)}$ is the ground state wavefunction width.  Under our conditions $\Gamma_{01} \ll 2\pi\times 1$ Hz~\cite{turchette2000}, indicating that cooling to $\bar{n} \ll 1$ should be feasible. 
The rate of heating in a real device, along with the actual dephasing rates for a molecule near the surface of a chip, are important phenomena which must be experimentally determined.

{\it Polar molecules as quantum bits: encoding, coherence properties, and one-bit gates}

Trapped polar molecules, cooled close to their ground state of motion in
the EZ trap and coupled to stripline resonators, represent a good starting
point for quantum bits.  For quantum processing, the qubit can be encoded
in a pair of trapped rotational states ($\ket{1}$ and $\ket{2}$) and
single qubit operations performed using classical microwave fields.
Molecules coupled through stripline resonators allow for two-qubit
operations.

We now consider coherence properties of rotational superpositions. Voltage fluctuations in the trap electrodes (quantified by $S_V(f)$) cause random fluctuations of the qubit transition frequency, and hence dephasing of rotational superposition states. Decoherence due to this voltage noise is determined by the field sensitivity of the rotational splitting, $\frac{\partial\omega_{0}}{\partial \mc{E}}$, and the effective RMS variations of the trap voltage, $V_{RMS}^{eff}$. With proper accounting of qubit phase fluctuations from a $1/f$ spectrum\cite{astafiev2004}, $V_{RMS}^{eff}   \sim 0.2 \ \mu$V.
In the linear Stark regime, and over a wide range in electric fields, the electric field sensitivity of the qubit transition frequency is  $\left|\frac{\partial\omega_{0}}{\partial \mc{E}}\right| \approx 0.1 \mu/\hbar~\approx 2\pi\times200$ kHz/(V/cm) for CaBr, leading to RMS frequency shifts $\delta\omega_0 = (\frac{\partial\omega_{0}}{\partial \mc{E}}) V_{RMS}/w$. For short times, this results in quadratic decay of qubit coherence\cite{schriefl2006}, with a characteristic rate $ \gamma_{V} \sim \delta\omega_0  = 2\pi \times 4 - 0.4$ kHz.

Although the trap potential is in general different for the two rotational qubit states, as assumed above, there exists a specific offset field value -- a ``sweet spot" -- for which the trap curvature is the same. (This occurs at $\mc{E}_{DC}^{\textrm{off, sweet}} \approx 3 \hbar B/\mu$.) When the trap is biased at this sweet spot,   
$\left|\frac{\partial\omega_{0}}{\partial \mc{E}}\right| = 0$ for the lowest vibrational levels of the trap.
The second order term is given by $\left|\frac{\partial^2\omega_{0}}{\partial \mc{E}^2}\right| \approx 0.1 \mu^2/(\hbar^2 B)$. In CaBr this is given numerically by $\Delta \omega_0 / \mc{E}_{DC}^2  = 2\pi \times 100 $ Hz (V cm$^{-1}$)$^{-2}$. Decoherence due to voltage noise can thus be vastly reduced by operating at the sweet spot, where $\delta \omega_0 \approx \left|\frac{\partial^2\omega_{0}}{\partial \mc{E}^2}\right|  \left(\frac{V_{RMS}^{eff}}{w}\right)^2$ and $\gamma_{V,2}\sim \delta \omega_0$  is below the 1 Hz level. Operation at the sweet spot comes with the modest expense of reducing the maximum possible trap depth by a factor of $4$.

Molecular motion also causes variations in the qubit level splitting, and
thus also  dephases rotational qubits.
If the molecule is in its motional ground state, ${\bar n} =0$, it can be
kept there during microwave manipulation of the qubit (provided the
excitation rate is slower than the trap frequency). In the case of finite
molecular temperature the rotational superpositions will dephase with
characteristic rate $\gamma_{T} \sim  (\omega_t^2/B) {\bar n}^2$. If
${\bar n} \sim 1$ and $\omega_t \sim 2\pi \times 5 $ MHz, then $\gamma_T
\sim 2\pi\times 1$ kHz. Thus, cooling of the molecular motion is crucial
for long-lived rotational coherences.

Single qubit quantum gates can be accomplished by driving rotational
transitions with oscillating electric fields. During such a gate operation (driven at Rabi frequency $\Omega$),
the error probability $p$ is bounded by $p \le (\gamma^*/\Omega)^2$ per
single Rabi cycle,  where $\gamma^*\equiv \gamma_{T}+\gamma_{V,2}$ is a
total dephasing rate for rotationally encoded qubits. Taking e.g. $\Omega \sim 2 \pi \times 1$MHz (consistent with all constraints on the system), we find that $p$ is
negligible for our operating conditions.

\begin{figure}[ht]
\begin{center}
\includegraphics[scale=0.1]{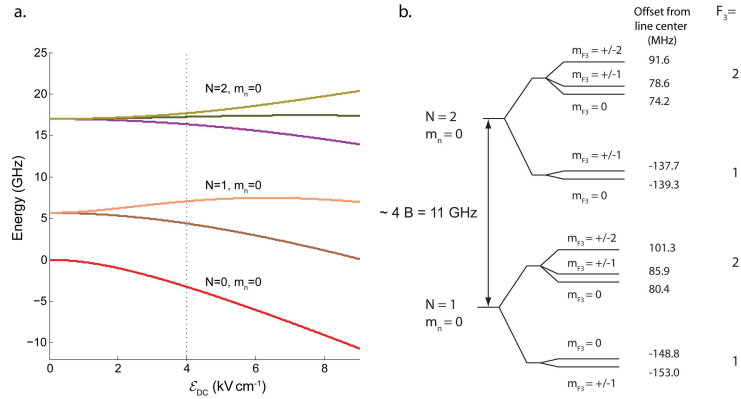}
\caption{
\label{fig:1}
{\bf Structure of selected rotational states of CaBr in an
electric field}}
\end{center}
\end{figure}

\begin{figure}[ht]
\begin{center}
\includegraphics[scale=.1]{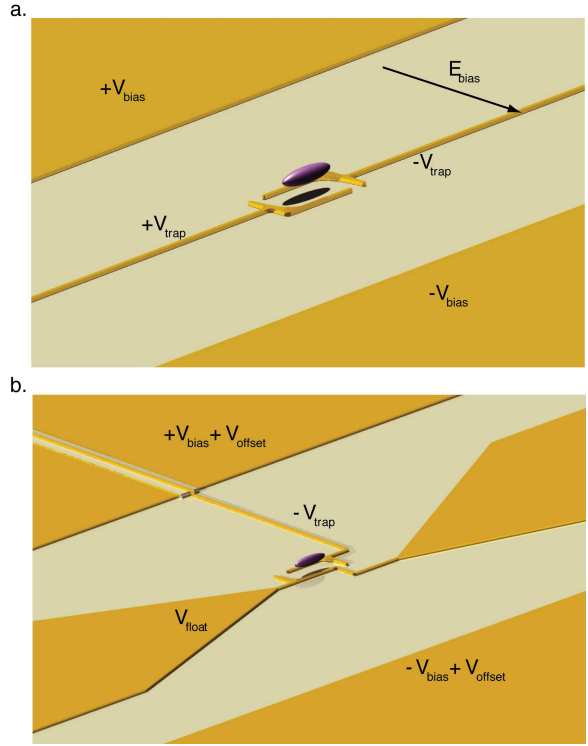}
\caption{
\label{fig:2}
{\bf Electrostatic Z-trap (EZ-trap).} 
}
\end{center}
\end{figure}

\begin{figure}
\begin{center}
\includegraphics[scale=0.1]{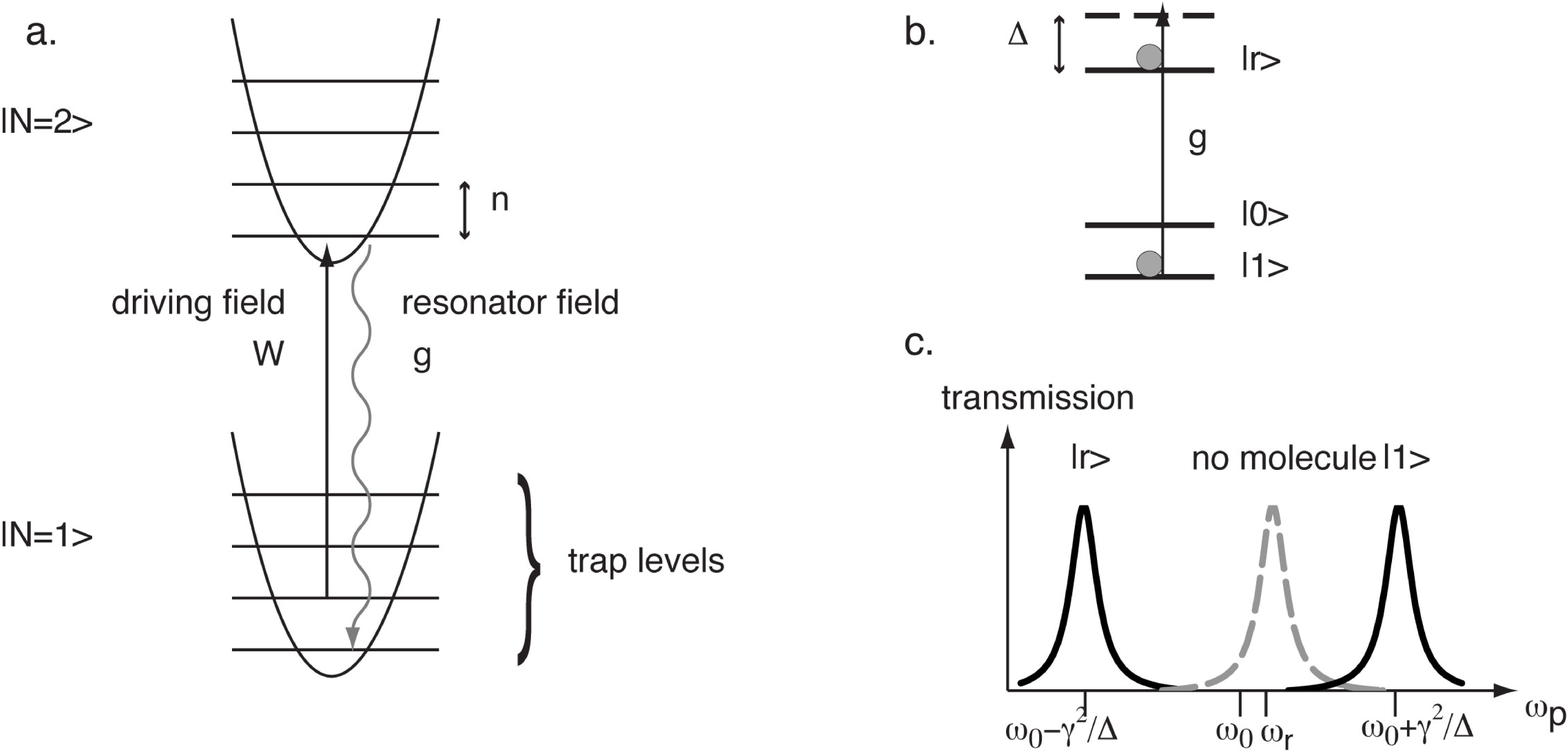}
\caption{
\label{fig:3}
{\bf Resonator-enhanced sideband cooling and quantum state readout.} 
\\
{\bf a.} Sideband cooling using resonator-enhanced spontaneous emission. The driving field is tuned to the red sideband $\ket{1,n}\rightarrow \ket{2,n-1}$, while the resonator is resonant with the $\ket{2,n}\rightarrow \ket{1,n}$ transition, where $n$ denotes the trap motional level.
{\bf b.} 
Quantum state readout via dispersive shift of cavity induced by the qubit. In the dispersive limit when the rotational transition of the molecule is significantly detuned from the cavity frequency ($\Delta_{r} \gg g$), a qubit state-dependent frequency shift $\delta \omega = \pm g^2/\Delta_{r}$ allows non-demolition measurement of the molecule's state by probing the transmission or reflection from the cavity. In the limit $\delta \omega < \kappa$, microwaves transmitted at the cavity frequency undergo a phase shift of $\pm\tan^{-1}\frac{2g^2}{\kappa\Delta_r}$ when the qubit is in state $\ket{1},\ket{2}$ respectively. 
{\bf c.} Probe field transmission versus probe frequency. When $g^2/\Delta_r>\kappa$, the frequency shift of the cavity is larger than the resonator linewidth. A probe beam at one or the other of the new, shifted frequencies will be transmitted or reflected, again allowing a potentially high-fidelity readout of the qubit state. In the absence of molecules, no frequency shift occurs, so the presence or absence of molecules in the trap can also be determined. }
\end{center}
\end{figure}

Finally, we note that for quantum storage the qubits can also be encoded
in hyperfine or Zeeman spin sublevels of a single rotational state, in
which low dephasing can be obtained even away from sweet spot. As
discussed below such encoding might facilitate scalability of the polar
molecule quantum computer. Our example case of CaBr has one nuclear spin
$I=3/2$ and one unpaired electron spin, with molecular state $^2\Sigma^+$.
Figure \ref{fig:1}a shows the energy levels of selected states in an
electric field. Within a broad range of electric field values, the value of
$\left|\frac{\partial\omega}{\partial \mc{E}}\right|$ for transitions
between hyperfine sublevels with $M_{F3} = 0$ is $\sim 10^3$ smaller
than for rotational transitions away from the
"sweet" spot.  Zeeman sublevels whose splittings are completely insensitive to
electric fields also exist (although these introduce new features beyond the scope of this paper). Note that hyperfine level degeneracies do not impose additional restrictions on one-bit gate speeds;  for CaBr near
the ``sweet spot",  the rotational transitions differ by $\delta \omega_{h}
\approx 15 $ MHz for two hyperfine states.

{\it Long-range quantum coupling of molecular qubits}

\begin{figure}
\begin{center}
\includegraphics[scale=0.1]{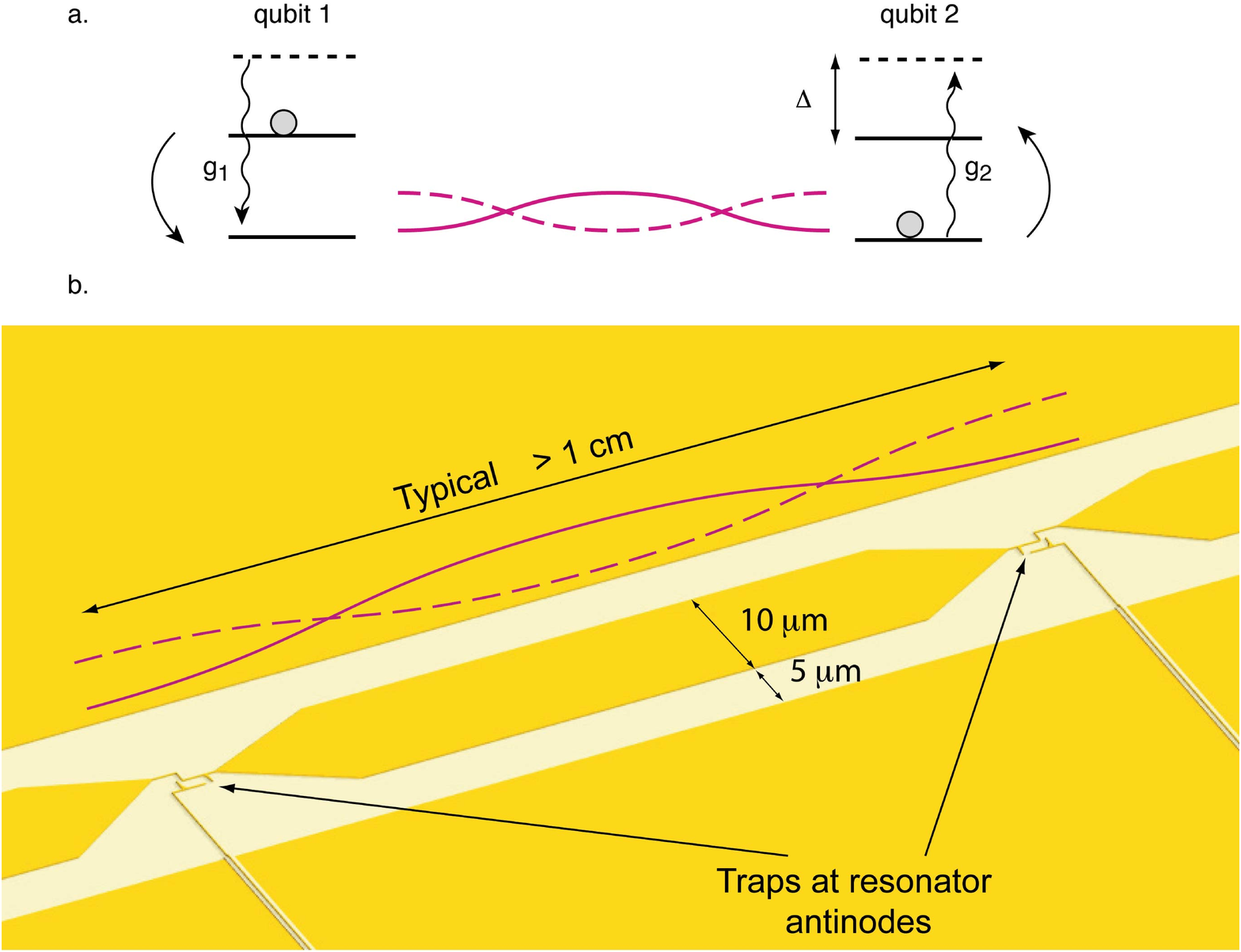}
\caption{
\label{fig:4}
{\bf Capacitive coupling of molecules mediated by stripline.} 
\\
{\bf a.} 
Polar molecule qubits are coupled to each other via (off-resonant) virtual exchange of microwave photons through a stripline resonator. The detuning between the resonator mode and the qubit frequency is $\Delta$, and the qubits are coupled to the resonator mode with the same vacuum Rabi frequencies $g$. The effective dipole-dipole interaction mediated by the resonator is given by the Hamiltonian 
$H_{int}=\frac{g^2}{\Delta}\left(\hat{\sigma}_{1}^{+}\hat{\sigma}_{2}^{-}+\hat{\sigma}_{1}^{-}\hat{\sigma}_{2}^{+}\right)$.
As indicated schematically on the figure, this interaction corresponds to qubit 1 emitting a virtual photon in the resonator while changing state from the upper to the lower state, and qubit 2 absorbing the virtual photon while changing state from the lower to the upper state.
{\bf b.} Multiple EZ traps can be patterned along the length of a stripline resonator, enabling coupling of multiple qubits. Here two EZ traps located at the resonator mode antinodes are shown, with typical dimensions as indicated on the figure.}

\end{center}
\end{figure}

We now consider coherent interaction of polar molecules through the capacitive, electrodynamic coupling to superconducting transmission line resonators \cite{sorensen:063601}. For simplicity, we consider a $\sqrt{\rm SWAP}$ operation between a pair of rotational qubits. 
As illustrated in Fig.~\ref{fig:4}, we use an off-resonant interaction with detuning $\Delta$ between resonator and qubit, with the molecules located at voltage nodes along the resonator. 
Assuming $\Delta \gg g$ and adiabatically eliminating the resonator degree of freedom, we find an interaction of the form $H_{int}=\frac{g^2}{\Delta}\left(\hat{\sigma}_{1}^+\hat{\sigma}_{2}^-+\hat{\sigma}_{1}^-\hat{\sigma}_{2}^+\right)$, where $\frac{g^2}{\Delta}$ is the interaction rate and
 $\hat{\sigma}_{i}^{-} = \ket{1}\bra{2}_i$ is the lowering operator for $i$th molecule.
 This effective exchange interaction can be used to map coherent superpositions from one quantum bit to another in time $\tau=\frac{\pi\Delta}{2g^2}$, thus enabling a universal two-qubit gate \cite{ciraczoller}. Note that at large molecule-resonator detuning ($\Delta \gg \kappa$), the resonator mode is only virtually occupied, so that cavity decay has little effect: the probability of error due to spontaneous emission of a photon during the two-bit gate is $p_{sp}=\kappa\frac{g^2}{\Delta^2}\tau$. While slower gate speed (at large detuning $\Delta$) results in reduced $p_{sp}$, it also results in increased probability of dephasing $p_{dep}=(\gamma^*\tau)^2$.
The overall probability of error $p_{err}=p_{sp}+p_{dep}$ is minimized by choosing
 $\Delta^*=\left(\frac{g^4\kappa}{\pi\gamma^{*2}}\right)^{1/3}$, resulting in a total error probability $p_{err} \approx \left(\frac{\kappa\gamma^*}{g^2}\right)^{2/3}$. For example, with $\kappa=2\pi\times 10$ kHz ($Q=10^6$), $g=2\pi\times 200$ kHz, and $\gamma^* \sim 1$ kHz we find that at the optimal detuning a probability of error is well below one percent. Thus, high-fidelity two-qubit operations between remote qubits are possible.

{\it State-dependent detection}

The presence of a trapped molecule, as well as its internal state, can be detected by measuring the phase shift of an off-resonant microwave field transmitted through the resonator. The dispersive interaction of the dipole and resonator leads to a state-dependent phase shift of the resonator field \cite{wallraff2004}, implementing a near-perfect quantum non-demolition (QND) measurement of the qubit state. 
This method has already been used to detect superconducting qubits \cite{wallraff:060501}, and it can also be applied to detecting the quantum state of the molecules with high signal to noise ratio. For example, with $Q=10^6$ and for a detuning of $\Delta_r=2\pi\times 5$ MHz, the phase shift of a microwave probe beam is $\theta_0 = {\rm tan}^{-1}[2 g^2/\kappa\Delta_r]\sim 10$ degrees. This phase shift can be measured using a probe beam with up to $n_{crit}=\Delta_r^2/4g^2 \sim 1000$ photons, or an incident power $P_{read} = n_{crit} \hbar\omega_0\kappa \sim 10^{-15\overline{}}$ Watts. The signal to noise ratio in one cavity lifetime is given by $SNR = {\rm sin}^2[\theta_0] (n_{crit}/n_{amp}) \sim 2$, where $n_{amp} = k_B T_N/\hbar\omega_0 \sim 20$ is the noise added by a cryogenic amplifier with a noise temperature of $T_N \sim 5$ K. During the readout, the rotational state of the molecule can still decay by spontaneous emission via the cavity at a rate $\gamma_\kappa=\kappa\frac{g^2}{\Delta_r^2}$, leading to a lifetime $T_1 = 1/\gamma_\kappa \sim 50$ ms; here the maximum SNR of the measurement is improved by a factor $\kappa/\gamma_\kappa \sim 5,000$, showing that very high fidelity ($> 99$\%) readout is feasible.

{\it Potential for scaling quantum circuits}

In order to realize scalable quantum information processors, or to
implement quantum error correction on logical qubits and high fidelity
quantum gates, the physical qubits need to be
connected to each other through quantum channels. In addition, quantum
information needs to be moved around the network efficiently. To implement
a logical qubit suitable for error correction, one requires the proximity
of several physical qubits, any pair of which can be subjected to a
quantum gate.
A possible solution involves an array of storage zones where molecules are
trapped, cooled and initialized via their coupling to stripline
resonators. Furthermore, in analogy to ion trap approaches \cite{kielpinksi2002},
qubits stored in hyperfine or Zeeman sublevels of a single rotational
state may be moved to interaction zones
using electrostatic gates. Moving and coupling of qubits could be achieved
by patterning multiple EZ traps in the gap between the two ground planes
of a stripline resonator.

In this scheme, we envision qubit transport in hyperfine/Zeeman states to relevant interaction zones and
processing in rotational states.
When a particular qubit is needed to perform a quantum operation, it can
be transferred from storage to a rotational state using frequency selective
microwave transitions. Taking the example of CaBr encoded in hyperfine
states, dephasing from either voltage noise or finite temperature of the
molecule is suppressed during storage and transport to very low levels. Specifically,
over a broad range of electric field values, a hyperfine qubit associated with
a motionally excited molecule will dephase at a rate
scaling as $\gamma_T^{hf} \sim \omega_t {\bar n}/10^3$.  This implies that for
reasonably cold molecules,  ${\bar n} = 1$, trap potentials can be changed
adiabatically (i.e. with negligible motional heating) on time
scale much faster than qubit dephasing.  The hyperfine qubit can be
transfered to rotational states when the molecule is brought back to the "sweet spot". Finally, we note that it may be possible to cool the
molecule's motion without destroying hyperfine/Zeeman state coherence,
provided that the detuning between cavity and qubit is
much larger than hyperfine/Zeeman level splitting.
Thus, potentially scalable quantum circuits could be designed using polar
molecules as quantum bits and superconducting resonators as the quantum
bus connecting these qubits.

{\it Outlook}

In this work, we have proposed an avenue in which the principal respective advantages of isolated atomic and mesoscopic solid state systems can be combined.  The resulting system provides tightly confined, coherent quantum systems with a high degree of control. Cooling and coherent manipulation of the quantum states of single molecules, as well as coherent coupling of molecules to one another, can be achieved near the surface of a superconducting chip using DC and microwave electric fields. We have shown that these techniques can be combined to yield a novel approach to quantum computation and that appropriately cooled molecules should have excellent coherence properties even in close proximity to the surface. Achievement of long
distance entanglement of molecular qubits via exchange of microwave
photons provides a complimentary approach to the shuttling of ions in
segmented traps, which underlies the scalability of the ion trap
quantum computer \cite{ciracphystoday2004}.

 A number of other interesting avenues can be considered. 
For example, the excellent coherence properties of hyperfine states of polar molecules may provide a long-term quantum memory for solid-state qubits. Specifically, 
coupling to a stripline resonator can be used to map the quantum state of a superconducting qubit \cite{wallraff:060501,makhlin2001,nakamura1999} to the state of either a single molecule, or perhaps also of a collective excitation of a small molecular ensemble through a collectively enhanced process\cite{rablnew}. 
In addition, this approach can be used to provide an interface to ``flying" qubits. Molecular microwave-frequency qubits 
can be mapped via a Raman process to single photons in the infrared or optical regime (corresponding to vibrational and electronic molecular transitions, respectively)\cite{lukin:457}.  In addition, novel approaches to controlled many-body physics can be envisaged. Tightly confined polar molecules in high aspect ratio EZ traps can be used to realize a one dimensional quantum system with strong, long-range interactions. Coupling to the stripline may provide a novel tool for preparation of such strongly interacting systems, as well as for detecting the resulting quantum phases \cite{sachdev}. 

We emphasize that the approach described here is unique in that
it combines tight localization, fast manipulation, and electrical gate
control, unprecedented for AMO systems, with the exceptional coherence
properties which are uncommon for condensed matter approaches. While the
techniques for production of cold polar molecules are not
developed as well as those for charged ions and neutral atoms, exciting
recent experimental progress indicates that the ideas proposed here are
within experimental reach.

\newpage
{\it Methods}

{\it Sideband cooling through enhanced spontaneous emission}

We assume the molecule is in a harmonic trap of frequency $\omega_t$, a distance $z$ above a stripline resonator of resonant frequency $\omega$, as shown in Fig.~\ref{fig:3}a. 
The resonator is tuned close to the resonance frequency $\omega_0$ of the molecular dipole transition $\ket{1}\rightarrow\ket{2}$. The height of the molecule above the resonator can be written as $z=z_0+\hat{z}$, where $\hat{z}$ is the displacement from equilibrium position in the trap.  We write $\hat{z}=a_0\hat{x}$.  Then $\hat{x}=(\hat{b}+\hat{b}^\dagger)$ is a dimensionless position operator, with $\hat{p}=-i(\hat{b}-\hat{b}^\dagger)$ its conjugate momentum, so that in the ground state $\Delta x =\Delta p =1$. 
Introducing the effective Lamb-Dicke parameter $\eta=a_0/z_0 \ll 1$ in analogy with sideband cooling in ion traps\cite{wineland79}, we have $g(\hat{x})=g_0\left[1-\eta\hat{x}+O[\eta^2]\right]$.
For our nominal parameters we find $a_0\simeq 3$ nm, so that for CaBr trapped at $z_0=100$ nm, $\eta \simeq 0.03$.

For cooling, the resonator is pumped by an external field  tuned to the red sideband, while the resonator field is resonant with the dipole transition, as shown in Fig.~\ref{fig:3}a. 
The coupling of the resonator field to the molecular dipole is described by the interaction Hamiltonian
\ba
\hat{H} &=&\omega_tb^+b^-
+g_0\left(\hat{a}^\dagger\sigma_-+\sigma_+\hat{a}\right)
\nonumber 
+ \eta g_0\hat{x}\left(\hat{a}^\dagger\sigma_-+\sigma_+\hat{a}\right).
\label{eq:delghamilt}
\ea
A general analysis based on perturbation theory using the small parameter $\eta$ can be used to obtain the cooling rates \cite{cirac1992}. We can also obtain approximate analytic expressions for the cooling rates in a simple regime of sideband cooling when $\omega_t\gg\kappa \gg g_0$. In this case the cavity field can be adiabatically eliminated, resulting in effective spontaneous decay of the excited rotational state at rate $\gamma_{sp} =2g^2/\kappa$. 
The sideband excitation rate $R$ from a coherent microwave drive with Rabi frequency $\Omega$ is given by $R= \eta^2 \Omega^2/\gamma_{sp}$.  
The cooling rate is then given by $\Gamma_c = \gamma_{sp}R/(2R+ \gamma_{sp})
\rightarrow \gamma_{sp}/2$ 
in the limit of strong driving. This rate is ultimately limited by cavity decay $\kappa/2$ when the strong coupling regime is reached. 

The resonator and trap degree of freedom exchange quanta until they equilibrate to the same mean number of quanta, which is set by the background (e.g., thermal) number of photons in the microwave cavity.
Note that while we describe here only cooling of the degree of freedom perpendicular to the chip surface, similar considerations can be applied to all trap degrees of freedom.
Finally, we note that the trapping potentials experienced by the molecule in the upper and lower states may be different. However, sideband cooling can still occur in this regime and leads to similar cooling rates and final temperatures as described above.

{\it Effect of surface on trapped molecules}

For a molecule at a small distance $z$ above a conducting surface ($z <
\omega/c$, where $\omega$ is the frequency of the dominant transition
contributing to the dipole moment),
there is a van der Waals (VDW) potential due to the attraction of the
molecule to its image dipole \cite{lin:050404}.
The potential is given approximately by
\begin{equation}
U(z) =\frac{\mu^2}{4 \pi \epsilon_0 (2 z)^3}=\frac{C_3}{z^3}\approx
\frac{20~[\mu\textrm{m}^3]}{z^3}\textrm{~nK},
\end{equation}
for CaBr.  Note that retardation does not modify the potential on the
micron scale, unlike in the case of atoms, because of the long wavelength of the
transition which contributes most to the dipole moment.
In addition, the trap frequency is modified by the $z^{-3}$ potential
and becomes
$\omega_t^\prime = \sqrt{\omega_t^2 - \frac{12 C_3}{ m z^5}}$ for a
harmonic potential centered at $z_0$.
At $z_0=100$~nm and $\omega_t=2\pi\times$6 MHz,
the change in the trapping frequency
$\Delta\omega_t/\omega_t \ll$ 1\% and there is no effect on the trap
depth due to the VDW interaction.
The trap depth begins to be affected for the case of smaller $z_0$ or
weaker confinement. For example, for $z_0=100$ nm and $\omega_t=2 \pi \times 1$ MHz, the van
der Waals potential shifts the trapping frequency by 2\% and reduces the
maximum trap depth from $\sim 3$~mK (set by the surface at z=0) to $< 1$~mK.

\newpage

{\bf Acknowledgments}

We thank T.Calarco, L Childress, A.Sorensen, and J.Taylor for helpful discussions.  Work at Harvard is supported by NSF, Harvard-MIT CUA
and Packard and Sloan Foundations. Work at Yale is
supported by NSF Grant DMR0325580, the W.M. Keck
Foundation, and the Army Research Office.
Work at Innsbruck is supported by the Austrian Science Foundation, European Networks and the Institute
for Quantum Information.

\newpage

{\bf Caption to Figure 1}
\\
Molecular structure and level shifts of CaBr relevant to the proposed schemes.\\
{\bf a.} Stark shifts of rotational levels in an applied electric field, showing the trappable states $\ket{1} \equiv \ket{N=1,m_N=0}$ and $\ket{2} \equiv \ket{N=2,m_N=0}$ (weak field seekers).  The dotted line marks the field value $\mathcal{E}_{DC}=\sweet$ for which the effective dipole moments of the weak field seeking states are the same.  Splittings due to electron and nuclear spin are too small to see on this scale.
\\
{\bf b.} Spin-rotation and hyperfine structure of Ca$^{79}$Br in a strong electric field. Energies shown are for $\mathcal{E}_{DC}=\sweet=4$~kV cm$^{-1}$.  The effects of electron and nuclear spin are determined by the spin-rotation Hamiltonian \cite{browncarrington} $H_{spin-rot}=\gamma_{sr}\bold{N}\cdot\bold{S}$ and the hyperfine Hamiltonian $H_{hfs}=b\bold{S}\cdot\bold{I}+c S_{z'}I_{z'}-e~\nabla\bold{E}\cdot\bold{Q}$, where $\hat{z}'$ is the molecular fixed internuclear axis, and the final term is the scalar product of two second rank tensors representing the gradient of the electric field at the location of the bromine nucleus and the electric quadrupole moment of that nucleus. For Ca$^{79}$Br, the size of the spin-rotation and hyperfine terms are comparable: $\gamma_{sr}=90.7$ MHz, $b=95.3$ MHz, $c=77.6$ MHz, and the electric quadrupole coupling constant $(eqQ)_0=20.0$~MHz\cite{childs}.  For large enough electric fields (such that $\mu \mc{E}_{DC} \gg \gamma_{sr}$), the nuclear spin $I$ and electron spin $S$ decouple from the rotational angular momentum $N$, while they couple to one another to form ${F}_3= S+I (= 1,2$ for CaBr).

{\bf Caption to Figure 2}
\\
{\bf a.} EZ-trap design. The thin wire-like electrodes biased at $\pm V_{trap}$ generate the strong local electric field gradients needed for trapping. A radial quadrupole field is created by the long sides of the electrodes, in combination with a transverse bias electric field $E_{bias}$ (created by the large electrodes with applied voltages $\pm V_{bias}$). This part is an electrostatic analog of the magnetic guides developed for ``atom chips''. Axial confinement is achieved by curving each electrode at the end and bringing it closer to the oppositely charged electrode, creating an increased electric field.  Like its magnetic counterpart, the trap is of the Ioffe-Pritchard type: there is no field zero, which avoids dipole flips from the field-aligned to the anti-aligned state (i.e. a ``Majorana hole'', which would enable coupling to the untrapped states with $m_N\neq 0$).  
\\
{\bf b.} Zoomed-out view of the EZ trap, integrated with a microwave stripline resonator. 
The ground planes of the resonator are biased at the DC voltages $\pm V_{bias}+V_{offset}$, giving rise to the bias field $E_{bias}$ for the EZ trap. The offset voltage $V_{offset}$ is used to bias the central pin and adjust $V_{float}$. In the region shown, which is of size much smaller than the wavelength of the microwave photons carried by the stripline, the width of the central pin of the stripline resonator is gradually reduced and deformed to the shape of one of the L-shaped electrodes of the EZ trap. The second L-shaped electrode necessary to form the EZ trap is made of a thin wire like electrode overlaid on one of the conducting ground planes. This electrode can behave as a continuation of the ground plane for AC voltages at microwave frequencies, while at DC it can be independently biased at the voltage $V_{trap}$, thereby completing the EZ trap. The overall effect of the region where the central pin is thinner is a slight change in the capacitance per unit length, without significantly affecting the quality of the resonator.

\pagebreak

\bibliography{paper_submit.bib}

\begin{thebibliography}{43}
\expandafter\ifx\csname natexlab\endcsname\relax\def\natexlab#1{#1}\fi
\expandafter\ifx\csname bibnamefont\endcsname\relax
  \def\bibnamefont#1{#1}\fi
\expandafter\ifx\csname bibfnamefont\endcsname\relax
  \def\bibfnamefont#1{#1}\fi
\expandafter\ifx\csname citenamefont\endcsname\relax
  \def\citenamefont#1{#1}\fi
\expandafter\ifx\csname url\endcsname\relax
  \def\url#1{\texttt{#1}}\fi
\expandafter\ifx\csname urlprefix\endcsname\relax\def\urlprefix{URL }\fi
\providecommand{\bibinfo}[2]{#2}
\providecommand{\eprint}[2][]{\url{#2}}

\bibitem[{\citenamefont{Doyle et~al.}(2004)\citenamefont{Doyle, Friedrich,
  Krems, and Masnou-Seeuws}}]{doyle2004}
\bibinfo{author}{\bibfnamefont{J.}~\bibnamefont{Doyle}},
  \bibinfo{author}{\bibfnamefont{B.}~\bibnamefont{Friedrich}},
  \bibinfo{author}{\bibfnamefont{R.~V.} \bibnamefont{Krems}}, \bibnamefont{and}
  \bibinfo{author}{\bibfnamefont{F.}~\bibnamefont{Masnou-Seeuws}},
  \bibinfo{journal}{European Physical Journal D} \textbf{\bibinfo{volume}{31}}
  (\bibinfo{year}{2004}).

\bibitem[{\citenamefont{Loss and DiVincenzo}(1998)}]{loss1998}
\bibinfo{author}{\bibfnamefont{D.}~\bibnamefont{Loss}} \bibnamefont{and}
  \bibinfo{author}{\bibfnamefont{D.}~\bibnamefont{DiVincenzo}},
  \bibinfo{journal}{Physical Review A} \textbf{\bibinfo{volume}{57}},
  \bibinfo{pages}{120Ð126} (\bibinfo{year}{1998}).

\bibitem[{\citenamefont{Wallraff et~al.}(2005)\citenamefont{Wallraff, Schuster,
  Blais, Frunzio, Majer, Devoret, Girvin, and Schoelkopf}}]{wallraff:060501}
\bibinfo{author}{\bibfnamefont{A.}~\bibnamefont{Wallraff}},
  \bibinfo{author}{\bibfnamefont{D.~I.} \bibnamefont{Schuster}},
  \bibinfo{author}{\bibfnamefont{A.}~\bibnamefont{Blais}},
  \bibinfo{author}{\bibfnamefont{L.}~\bibnamefont{Frunzio}},
  \bibinfo{author}{\bibfnamefont{J.}~\bibnamefont{Majer}},
  \bibinfo{author}{\bibfnamefont{M.~H.} \bibnamefont{Devoret}},
  \bibinfo{author}{\bibfnamefont{S.~M.} \bibnamefont{Girvin}},
  \bibnamefont{and} \bibinfo{author}{\bibfnamefont{R.~J.}
  \bibnamefont{Schoelkopf}}, \bibinfo{journal}{Physical Review Letters}
  \textbf{\bibinfo{volume}{95}}, \bibinfo{eid}{060501} (\bibinfo{year}{2005}).

\bibitem[{\citenamefont{Makhlin et~al.}(2001)\citenamefont{Makhlin, Sch\"{o}n,
  and Shnirmam}}]{makhlin2001}
\bibinfo{author}{\bibfnamefont{Y.}~\bibnamefont{Makhlin}},
  \bibinfo{author}{\bibfnamefont{G.}~\bibnamefont{Sch\"{o}n}},
  \bibnamefont{and} \bibinfo{author}{\bibfnamefont{A.}~\bibnamefont{Shnirmam}},
  \bibinfo{journal}{Reviews of Modern Physics} \textbf{\bibinfo{volume}{73}},
  \bibinfo{pages}{357Ð400} (\bibinfo{year}{2001}).

\bibitem[{\citenamefont{Nakamura et~al.}(1999)\citenamefont{Nakamura, Pashkin,
  and Tsai}}]{nakamura1999}
\bibinfo{author}{\bibfnamefont{Y.}~\bibnamefont{Nakamura}},
  \bibinfo{author}{\bibfnamefont{Y.}~\bibnamefont{Pashkin}}, \bibnamefont{and}
  \bibinfo{author}{\bibfnamefont{J.}~\bibnamefont{Tsai}},
  \bibinfo{journal}{Nature} \textbf{\bibinfo{volume}{398}},
  \bibinfo{pages}{786Ð788} (\bibinfo{year}{1999}).

\bibitem[{\citenamefont{Mandel et~al.}(2003)\citenamefont{Mandel, Greiner,
  Widera, Rom, H\"{a}nsch, and Bloch}}]{mandel2003}
\bibinfo{author}{\bibfnamefont{O.}~\bibnamefont{Mandel}},
  \bibinfo{author}{\bibfnamefont{M.}~\bibnamefont{Greiner}},
  \bibinfo{author}{\bibfnamefont{A.}~\bibnamefont{Widera}},
  \bibinfo{author}{\bibfnamefont{T.}~\bibnamefont{Rom}},
  \bibinfo{author}{\bibfnamefont{T.}~\bibnamefont{H\"{a}nsch}},
  \bibnamefont{and} \bibinfo{author}{\bibfnamefont{I.}~\bibnamefont{Bloch}},
  \bibinfo{journal}{Nature} \textbf{\bibinfo{volume}{425}},
  \bibinfo{pages}{937} (\bibinfo{year}{2003}).

\bibitem[{\citenamefont{Folman et~al.}(2002)\citenamefont{Folman, Kr\"{u}ger,
  Schmiedmayer, Denschlag, and Henkel}}]{folman2002}
\bibinfo{author}{\bibfnamefont{R.}~\bibnamefont{Folman}},
  \bibinfo{author}{\bibfnamefont{P.}~\bibnamefont{Kr\"{u}ger}},
  \bibinfo{author}{\bibfnamefont{J.}~\bibnamefont{Schmiedmayer}},
  \bibinfo{author}{\bibfnamefont{J.}~\bibnamefont{Denschlag}},
  \bibnamefont{and} \bibinfo{author}{\bibfnamefont{C.}~\bibnamefont{Henkel}},
  \bibinfo{journal}{Advances in Atomic, Molecular, and Optical Physics}
  \textbf{\bibinfo{volume}{48}}, \bibinfo{pages}{263} (\bibinfo{year}{2002}).

\bibitem[{\citenamefont{Cirac and Zoller}(2004)}]{ciracphystoday2004}
\bibinfo{author}{\bibfnamefont{J.~I.} \bibnamefont{Cirac}} \bibnamefont{and}
  \bibinfo{author}{\bibfnamefont{P.}~\bibnamefont{Zoller}},
  \bibinfo{journal}{Physics Today} \textbf{\bibinfo{volume}{57}},
  \bibinfo{pages}{38} (\bibinfo{year}{2004}).

\bibitem[{\citenamefont{Leibfried et~al.}(2005)\citenamefont{Leibfried, Knill,
  Seidelin, Britton, Blakestad, Chiaverini, Hume, Itano, Jost, Langer
  et~al.}}]{leibfried2005}
\bibinfo{author}{\bibfnamefont{D.}~\bibnamefont{Leibfried}},
  \bibinfo{author}{\bibfnamefont{E.}~\bibnamefont{Knill}},
  \bibinfo{author}{\bibfnamefont{S.}~\bibnamefont{Seidelin}},
  \bibinfo{author}{\bibfnamefont{J.}~\bibnamefont{Britton}},
  \bibinfo{author}{\bibfnamefont{R.~B.} \bibnamefont{Blakestad}},
  \bibinfo{author}{\bibfnamefont{J.}~\bibnamefont{Chiaverini}},
  \bibinfo{author}{\bibfnamefont{D.~B.} \bibnamefont{Hume}},
  \bibinfo{author}{\bibfnamefont{W.~M.} \bibnamefont{Itano}},
  \bibinfo{author}{\bibfnamefont{J.~D.} \bibnamefont{Jost}},
  \bibinfo{author}{\bibfnamefont{C.}~\bibnamefont{Langer}},
  \bibnamefont{et~al.}, \bibinfo{journal}{Nature}
  \textbf{\bibinfo{volume}{438}}, \bibinfo{pages}{639} (\bibinfo{year}{2005}).

\bibitem[{\citenamefont{H\"{a}ffner et~al.}(2005)\citenamefont{H\"{a}ffner,
  H\"{a}nsel, Roos, Benhelm, al~kar, Chwalla, K\"{o}rber, Rapol, Riebe, Schmidt
  et~al.}}]{haffner2005}
\bibinfo{author}{\bibfnamefont{H.}~\bibnamefont{H\"{a}ffner}},
  \bibinfo{author}{\bibfnamefont{W.}~\bibnamefont{H\"{a}nsel}},
  \bibinfo{author}{\bibfnamefont{C.~F.} \bibnamefont{Roos}},
  \bibinfo{author}{\bibfnamefont{J.}~\bibnamefont{Benhelm}},
  \bibinfo{author}{\bibfnamefont{D.~C.} \bibnamefont{al~kar}},
  \bibinfo{author}{\bibfnamefont{M.}~\bibnamefont{Chwalla}},
  \bibinfo{author}{\bibfnamefont{T.}~\bibnamefont{K\"{o}rber}},
  \bibinfo{author}{\bibfnamefont{U.~D.} \bibnamefont{Rapol}},
  \bibinfo{author}{\bibfnamefont{M.}~\bibnamefont{Riebe}},
  \bibinfo{author}{\bibfnamefont{P.~O.} \bibnamefont{Schmidt}},
  \bibnamefont{et~al.}, \bibinfo{journal}{Nature}
  \textbf{\bibinfo{volume}{438}}, \bibinfo{pages}{643} (\bibinfo{year}{2005}).

\bibitem[{\citenamefont{DeMille}(2002)}]{demille:067901}
\bibinfo{author}{\bibfnamefont{D.}~\bibnamefont{DeMille}},
  \bibinfo{journal}{Physical Review Letters} \textbf{\bibinfo{volume}{88}},
  \bibinfo{pages}{067901} (\bibinfo{year}{2002}).

\bibitem[{\citenamefont{Bethlem et~al.}(2000)\citenamefont{Bethlem, Berden,
  Crompvoets, Jongma, van Roij, and Meijer}}]{bethlem2000}
\bibinfo{author}{\bibfnamefont{H.~L.} \bibnamefont{Bethlem}},
  \bibinfo{author}{\bibfnamefont{G.}~\bibnamefont{Berden}},
  \bibinfo{author}{\bibfnamefont{F.~M.~H.} \bibnamefont{Crompvoets}},
  \bibinfo{author}{\bibfnamefont{R.~T.} \bibnamefont{Jongma}},
  \bibinfo{author}{\bibfnamefont{A.~J.~A.} \bibnamefont{van Roij}},
  \bibnamefont{and} \bibinfo{author}{\bibfnamefont{G.}~\bibnamefont{Meijer}},
  \bibinfo{journal}{Nature} \textbf{\bibinfo{volume}{406}},
  \bibinfo{pages}{491} (\bibinfo{year}{2000}).

\bibitem[{\citenamefont{Rieger et~al.}(2005)\citenamefont{Rieger, Junglen,
  Rangwala, Pinkse, and Rempe}}]{rieger:173002}
\bibinfo{author}{\bibfnamefont{T.}~\bibnamefont{Rieger}},
  \bibinfo{author}{\bibfnamefont{T.}~\bibnamefont{Junglen}},
  \bibinfo{author}{\bibfnamefont{S.~A.} \bibnamefont{Rangwala}},
  \bibinfo{author}{\bibfnamefont{P.~W.~H.} \bibnamefont{Pinkse}},
  \bibnamefont{and} \bibinfo{author}{\bibfnamefont{G.}~\bibnamefont{Rempe}},
  \bibinfo{journal}{Physical Review Letters} \textbf{\bibinfo{volume}{95}},
  \bibinfo{eid}{173002} (\bibinfo{year}{2005}).

\bibitem[{\citenamefont{Xia et~al.}(2005)\citenamefont{Xia, Deng, and
  Yin}}]{xia2005}
\bibinfo{author}{\bibfnamefont{Y.}~\bibnamefont{Xia}},
  \bibinfo{author}{\bibfnamefont{L.}~\bibnamefont{Deng}}, \bibnamefont{and}
  \bibinfo{author}{\bibfnamefont{J.}~\bibnamefont{Yin}},
  \bibinfo{journal}{Applied Physics B} \textbf{\bibinfo{volume}{81}},
  \bibinfo{pages}{459} (\bibinfo{year}{2005}).

\bibitem[{\citenamefont{Egorov et~al.}(2002)\citenamefont{Egorov, Lahaye,
  Schollkopf, Friedrich, and Doyle}}]{egorov:043401}
\bibinfo{author}{\bibfnamefont{D.}~\bibnamefont{Egorov}},
  \bibinfo{author}{\bibfnamefont{T.}~\bibnamefont{Lahaye}},
  \bibinfo{author}{\bibfnamefont{W.}~\bibnamefont{Schollkopf}},
  \bibinfo{author}{\bibfnamefont{B.}~\bibnamefont{Friedrich}},
  \bibnamefont{and} \bibinfo{author}{\bibfnamefont{J.~M.} \bibnamefont{Doyle}},
  \bibinfo{journal}{Physical Review A} \textbf{\bibinfo{volume}{66}},
  \bibinfo{eid}{043401} (\bibinfo{year}{2002}).

\bibitem[{\citenamefont{Weinstein et~al.}(2002)\citenamefont{Weinstein,
  deCarvalho, Hancox, and Doyle}}]{weinstein2002}
\bibinfo{author}{\bibfnamefont{J.~D.} \bibnamefont{Weinstein}},
  \bibinfo{author}{\bibfnamefont{R.}~\bibnamefont{deCarvalho}},
  \bibinfo{author}{\bibfnamefont{C.~I.} \bibnamefont{Hancox}},
  \bibnamefont{and} \bibinfo{author}{\bibfnamefont{J.~M.} \bibnamefont{Doyle}},
  \bibinfo{journal}{Physical Review A} \textbf{\bibinfo{volume}{65}},
  \bibinfo{pages}{021604(R)} (\bibinfo{year}{2002}).

\bibitem[{\citenamefont{DeMille et~al.}(2004)\citenamefont{DeMille, Glenn, and
  Petricka}}]{demille04}
\bibinfo{author}{\bibfnamefont{D.}~\bibnamefont{DeMille}},
  \bibinfo{author}{\bibfnamefont{D.}~\bibnamefont{Glenn}}, \bibnamefont{and}
  \bibinfo{author}{\bibfnamefont{J.}~\bibnamefont{Petricka}},
  \bibinfo{journal}{European Physical Journal D} \textbf{\bibinfo{volume}{31}},
  \bibinfo{pages}{375} (\bibinfo{year}{2004}).

\bibitem[{\citenamefont{Masuhara et~al.}(1988)\citenamefont{Masuhara, Doyle,
  Sandberg, Kleppner, Greytak, Hess, and Kochanski}}]{masuhara88}
\bibinfo{author}{\bibfnamefont{N.}~\bibnamefont{Masuhara}},
  \bibinfo{author}{\bibfnamefont{J.~M.} \bibnamefont{Doyle}},
  \bibinfo{author}{\bibfnamefont{J.~C.} \bibnamefont{Sandberg}},
  \bibinfo{author}{\bibfnamefont{D.}~\bibnamefont{Kleppner}},
  \bibinfo{author}{\bibfnamefont{T.~J.} \bibnamefont{Greytak}},
  \bibinfo{author}{\bibfnamefont{H.~F.} \bibnamefont{Hess}}, \bibnamefont{and}
  \bibinfo{author}{\bibfnamefont{G.~P.} \bibnamefont{Kochanski}},
  \bibinfo{journal}{Physical Review Letters} \textbf{\bibinfo{volume}{61}},
  \bibinfo{pages}{934} (\bibinfo{year}{1988}).

\bibitem[{\citenamefont{Vuletic et~al.}(2001)\citenamefont{Vuletic, Chan, and
  Black}}]{vuletic:033405}
\bibinfo{author}{\bibfnamefont{V.}~\bibnamefont{Vuletic}},
  \bibinfo{author}{\bibfnamefont{H.~W.} \bibnamefont{Chan}}, \bibnamefont{and}
  \bibinfo{author}{\bibfnamefont{A.~T.} \bibnamefont{Black}},
  \bibinfo{journal}{Physical Review A (Atomic, Molecular, and Optical Physics)}
  \textbf{\bibinfo{volume}{64}}, \bibinfo{eid}{033405} (\bibinfo{year}{2001}).

\bibitem[{\citenamefont{Boxhinski et~al.}(2003)\citenamefont{Boxhinski, Hudson,
  Lewandowski, Meijer, and Ye}}]{bochinski2003}
\bibinfo{author}{\bibfnamefont{J.}~\bibnamefont{Boxhinski}},
  \bibinfo{author}{\bibfnamefont{E.~R.} \bibnamefont{Hudson}},
  \bibinfo{author}{\bibfnamefont{H.}~\bibnamefont{Lewandowski}},
  \bibinfo{author}{\bibfnamefont{G.}~\bibnamefont{Meijer}}, \bibnamefont{and}
  \bibinfo{author}{\bibfnamefont{J.}~\bibnamefont{Ye}},
  \bibinfo{journal}{Physical Review Letters} \textbf{\bibinfo{volume}{91}},
  \bibinfo{pages}{243001} (\bibinfo{year}{2003}).

\bibitem[{\citenamefont{Wallraff et~al.}(2004)\citenamefont{Wallraff, Schuster,
  Blais, Frunzio, Huang, Majer, Kumar, Girvin, and Schoelkopf}}]{wallraff2004}
\bibinfo{author}{\bibfnamefont{A.}~\bibnamefont{Wallraff}},
  \bibinfo{author}{\bibfnamefont{D.~I.} \bibnamefont{Schuster}},
  \bibinfo{author}{\bibfnamefont{A.}~\bibnamefont{Blais}},
  \bibinfo{author}{\bibfnamefont{L.}~\bibnamefont{Frunzio}},
  \bibinfo{author}{\bibfnamefont{R.~S.} \bibnamefont{Huang}},
  \bibinfo{author}{\bibfnamefont{J.}~\bibnamefont{Majer}},
  \bibinfo{author}{\bibfnamefont{S.}~\bibnamefont{Kumar}},
  \bibinfo{author}{\bibfnamefont{S.~M.} \bibnamefont{Girvin}},
  \bibnamefont{and} \bibinfo{author}{\bibfnamefont{R.~J.}
  \bibnamefont{Schoelkopf}}, \bibinfo{journal}{Nature}
  \textbf{\bibinfo{volume}{431}}, \bibinfo{pages}{162} (\bibinfo{year}{2004}).

\bibitem[{\citenamefont{Blais et~al.}(2004)\citenamefont{Blais, Huang,
  Wallraff, Girvin, and Schoelkopf}}]{blais:062320}
\bibinfo{author}{\bibfnamefont{A.}~\bibnamefont{Blais}},
  \bibinfo{author}{\bibfnamefont{R.-S.} \bibnamefont{Huang}},
  \bibinfo{author}{\bibfnamefont{A.}~\bibnamefont{Wallraff}},
  \bibinfo{author}{\bibfnamefont{S.~M.} \bibnamefont{Girvin}},
  \bibnamefont{and} \bibinfo{author}{\bibfnamefont{R.~J.}
  \bibnamefont{Schoelkopf}}, \bibinfo{journal}{Physical Review A}
  \textbf{\bibinfo{volume}{69}}, \bibinfo{pages}{062320}
  (\bibinfo{year}{2004}).

\bibitem[{\citenamefont{Scully and Zubairy}(1997)}]{scullyQO}
\bibinfo{author}{\bibfnamefont{M.}~\bibnamefont{Scully}} \bibnamefont{and}
  \bibinfo{author}{\bibfnamefont{M.~S.} \bibnamefont{Zubairy}},
  \emph{\bibinfo{title}{Quantum Optics}} (\bibinfo{publisher}{Cambridge
  University Press}, \bibinfo{address}{Cambridge}, \bibinfo{year}{1997}).

\bibitem[{\citenamefont{Sorensen et~al.}(2004)\citenamefont{Sorensen, van~der
  Wal, Childress, and Lukin}}]{sorensen:063601}
\bibinfo{author}{\bibfnamefont{A.~S.} \bibnamefont{Sorensen}},
  \bibinfo{author}{\bibfnamefont{C.~H.} \bibnamefont{van~der Wal}},
  \bibinfo{author}{\bibfnamefont{L.~I.} \bibnamefont{Childress}},
  \bibnamefont{and} \bibinfo{author}{\bibfnamefont{M.~D.} \bibnamefont{Lukin}},
  \bibinfo{journal}{Physical Review Letters} \textbf{\bibinfo{volume}{92}},
  \bibinfo{pages}{063601} (\bibinfo{year}{2004}).

\bibitem[{\citenamefont{Wallace and Silsbee}(1991)}]{wallace:1754}
\bibinfo{author}{\bibfnamefont{W.~J.} \bibnamefont{Wallace}} \bibnamefont{and}
  \bibinfo{author}{\bibfnamefont{R.~H.} \bibnamefont{Silsbee}},
  \bibinfo{journal}{Review of Scientific Instruments}
  \textbf{\bibinfo{volume}{62}}, \bibinfo{pages}{1754} (\bibinfo{year}{1991}).

\bibitem[{\citenamefont{Frunzio et~al.}(2005)\citenamefont{Frunzio, Wallraff,
  Schuster, Majer, and Schoelkopf}}]{frunzio}
\bibinfo{author}{\bibfnamefont{L.}~\bibnamefont{Frunzio}},
  \bibinfo{author}{\bibfnamefont{A.}~\bibnamefont{Wallraff}},
  \bibinfo{author}{\bibfnamefont{D.}~\bibnamefont{Schuster}},
  \bibinfo{author}{\bibfnamefont{J.}~\bibnamefont{Majer}}, \bibnamefont{and}
  \bibinfo{author}{\bibfnamefont{R.}~\bibnamefont{Schoelkopf}},
  \bibinfo{journal}{IEEE Trans. Appl. Supercond.}
  \textbf{\bibinfo{volume}{15}}, \bibinfo{pages}{860} (\bibinfo{year}{2005}).

\bibitem[{\citenamefont{Raimond et~al.}(2005)\citenamefont{Raimond, Meunier,
  Bertet, Gleyzes, Maioli, Auffeves, Nogues, Brune, and Haroche}}]{raimond2005}
\bibinfo{author}{\bibfnamefont{J.~M.} \bibnamefont{Raimond}},
  \bibinfo{author}{\bibfnamefont{T.}~\bibnamefont{Meunier}},
  \bibinfo{author}{\bibfnamefont{P.}~\bibnamefont{Bertet}},
  \bibinfo{author}{\bibfnamefont{S.}~\bibnamefont{Gleyzes}},
  \bibinfo{author}{\bibfnamefont{P.}~\bibnamefont{Maioli}},
  \bibinfo{author}{\bibfnamefont{A.}~\bibnamefont{Auffeves}},
  \bibinfo{author}{\bibfnamefont{G.}~\bibnamefont{Nogues}},
  \bibinfo{author}{\bibfnamefont{M.}~\bibnamefont{Brune}}, \bibnamefont{and}
  \bibinfo{author}{\bibfnamefont{S.}~\bibnamefont{Haroche}},
  \bibinfo{journal}{Journal of Physics B} \textbf{\bibinfo{volume}{38}},
  \bibinfo{pages}{S535} (\bibinfo{year}{2005}).

\bibitem[{\citenamefont{Miller et~al.}(2005)\citenamefont{Miller, Northup,
  Birnbaum, Boca, Boozer, and Kimble}}]{miller2005}
\bibinfo{author}{\bibfnamefont{R.}~\bibnamefont{Miller}},
  \bibinfo{author}{\bibfnamefont{T.~E.} \bibnamefont{Northup}},
  \bibinfo{author}{\bibfnamefont{K.~M.} \bibnamefont{Birnbaum}},
  \bibinfo{author}{\bibfnamefont{A.}~\bibnamefont{Boca}},
  \bibinfo{author}{\bibfnamefont{A.~D.} \bibnamefont{Boozer}},
  \bibnamefont{and} \bibinfo{author}{\bibfnamefont{H.~J.}
  \bibnamefont{Kimble}}, \bibinfo{journal}{Journal of Physics B}
  \textbf{\bibinfo{volume}{38}}, \bibinfo{pages}{S551} (\bibinfo{year}{2005}).

\bibitem[{\citenamefont{Wineland and Itano}(1979)}]{wineland79}
\bibinfo{author}{\bibfnamefont{D.~J.} \bibnamefont{Wineland}} \bibnamefont{and}
  \bibinfo{author}{\bibfnamefont{W.~M.} \bibnamefont{Itano}},
  \bibinfo{journal}{Physical Review A} \textbf{\bibinfo{volume}{20}},
  \bibinfo{pages}{1521} (\bibinfo{year}{1979}).

\bibitem[{\citenamefont{Deslauriers et~al.}(2004)\citenamefont{Deslauriers,
  Haljan, Lee, Brickman, Blinov, Madsen, and Monroe}}]{deslauriers:043408}
\bibinfo{author}{\bibfnamefont{L.}~\bibnamefont{Deslauriers}},
  \bibinfo{author}{\bibfnamefont{P.~C.} \bibnamefont{Haljan}},
  \bibinfo{author}{\bibfnamefont{P.~J.} \bibnamefont{Lee}},
  \bibinfo{author}{\bibfnamefont{K.-A.} \bibnamefont{Brickman}},
  \bibinfo{author}{\bibfnamefont{B.~B.} \bibnamefont{Blinov}},
  \bibinfo{author}{\bibfnamefont{M.~J.} \bibnamefont{Madsen}},
  \bibnamefont{and} \bibinfo{author}{\bibfnamefont{C.}~\bibnamefont{Monroe}},
  \bibinfo{journal}{Physical Review A} \textbf{\bibinfo{volume}{70}},
  \bibinfo{pages}{043408} (\bibinfo{year}{2004}).

\bibitem[{\citenamefont{Zorin et~al.}(1996)\citenamefont{Zorin, Ahlers,
  Niemeyer, Weimann, Wolf, Krupenin, and Lotkhov}}]{Zorin96}
\bibinfo{author}{\bibfnamefont{A.}~\bibnamefont{Zorin}},
  \bibinfo{author}{\bibfnamefont{F.~J.} \bibnamefont{Ahlers}},
  \bibinfo{author}{\bibfnamefont{J.}~\bibnamefont{Niemeyer}},
  \bibinfo{author}{\bibfnamefont{T.}~\bibnamefont{Weimann}},
  \bibinfo{author}{\bibfnamefont{H.}~\bibnamefont{Wolf}},
  \bibinfo{author}{\bibfnamefont{V.~A.} \bibnamefont{Krupenin}},
  \bibnamefont{and} \bibinfo{author}{\bibfnamefont{S.~V.}
  \bibnamefont{Lotkhov}}, \bibinfo{journal}{Physical Review B}
  \textbf{\bibinfo{volume}{53}}, \bibinfo{pages}{13682} (\bibinfo{year}{1996}).

\bibitem[{\citenamefont{Turchette et~al.}(2000)\citenamefont{Turchette,
  Kielpinski, King, Leibfried, Meekhof, Myatt, Rowe, Sackett, Wood, Itano
  et~al.}}]{turchette2000}
\bibinfo{author}{\bibfnamefont{Q.~A.} \bibnamefont{Turchette}},
  \bibinfo{author}{\bibfnamefont{D.}~\bibnamefont{Kielpinski}},
  \bibinfo{author}{\bibfnamefont{B.~E.} \bibnamefont{King}},
  \bibinfo{author}{\bibfnamefont{D.}~\bibnamefont{Leibfried}},
  \bibinfo{author}{\bibfnamefont{D.~M.} \bibnamefont{Meekhof}},
  \bibinfo{author}{\bibfnamefont{C.~J.} \bibnamefont{Myatt}},
  \bibinfo{author}{\bibfnamefont{M.~A.} \bibnamefont{Rowe}},
  \bibinfo{author}{\bibfnamefont{C.~A.} \bibnamefont{Sackett}},
  \bibinfo{author}{\bibfnamefont{C.~S.} \bibnamefont{Wood}},
  \bibinfo{author}{\bibfnamefont{W.~M.} \bibnamefont{Itano}},
  \bibnamefont{et~al.}, \bibinfo{journal}{Physical Review A}
  \textbf{\bibinfo{volume}{61}}, \bibinfo{pages}{063418}
  (\bibinfo{year}{2000}).

\bibitem[{\citenamefont{Astafiev et~al.}(2004)\citenamefont{Astafiev, Pashkin,
  Nakamura, Yamamoto, and Tsai}}]{astafiev2004}
\bibinfo{author}{\bibfnamefont{O.}~\bibnamefont{Astafiev}},
  \bibinfo{author}{\bibfnamefont{Y.~A.} \bibnamefont{Pashkin}},
  \bibinfo{author}{\bibfnamefont{Y.}~\bibnamefont{Nakamura}},
  \bibinfo{author}{\bibfnamefont{T.}~\bibnamefont{Yamamoto}}, \bibnamefont{and}
  \bibinfo{author}{\bibfnamefont{J.}~\bibnamefont{Tsai}},
  \bibinfo{journal}{Physical Review Letters} \textbf{\bibinfo{volume}{93}},
  \bibinfo{pages}{267007} (\bibinfo{year}{2004}).

\bibitem[{\citenamefont{Schriefl et~al.}(2006)\citenamefont{Schriefl, Makhlin,
  Shnirman, and Schoen}}]{schriefl2006}
\bibinfo{author}{\bibfnamefont{J.}~\bibnamefont{Schriefl}},
  \bibinfo{author}{\bibfnamefont{Y.}~\bibnamefont{Makhlin}},
  \bibinfo{author}{\bibfnamefont{A.}~\bibnamefont{Shnirman}}, \bibnamefont{and}
  \bibinfo{author}{\bibfnamefont{G.}~\bibnamefont{Schoen}},
  \bibinfo{journal}{New Journal of Physics} \textbf{\bibinfo{volume}{8}},
  \bibinfo{pages}{001} (\bibinfo{year}{2006}).

\bibitem[{\citenamefont{Cirac and Zoller}(1995)}]{ciraczoller}
\bibinfo{author}{\bibfnamefont{J.~I.} \bibnamefont{Cirac}} \bibnamefont{and}
  \bibinfo{author}{\bibfnamefont{P.}~\bibnamefont{Zoller}},
  \bibinfo{journal}{Physical Review Letters} \textbf{\bibinfo{volume}{74}},
  \bibinfo{pages}{4091} (\bibinfo{year}{1995}).

\bibitem[{\citenamefont{Kielpinski et~al.}(2002)\citenamefont{Kielpinski,
  Monroe, and Wineland}}]{kielpinksi2002}
\bibinfo{author}{\bibfnamefont{D.}~\bibnamefont{Kielpinski}},
  \bibinfo{author}{\bibfnamefont{C.}~\bibnamefont{Monroe}}, \bibnamefont{and}
  \bibinfo{author}{\bibfnamefont{D.~J.} \bibnamefont{Wineland}},
  \bibinfo{journal}{Nature} \textbf{\bibinfo{volume}{417}},
  \bibinfo{pages}{709} (\bibinfo{year}{2002}).

\bibitem[{\citenamefont{Rabl et~al.}(2006)\citenamefont{Rabl, DeMille, Doyle,
  Lukin, Schoelkopf, and Zoller}}]{rablnew}
\bibinfo{author}{\bibfnamefont{P.}~\bibnamefont{Rabl}},
  \bibinfo{author}{\bibfnamefont{D.}~\bibnamefont{DeMille}},
  \bibinfo{author}{\bibfnamefont{J.~M.} \bibnamefont{Doyle}},
  \bibinfo{author}{\bibfnamefont{M.~D.} \bibnamefont{Lukin}},
  \bibinfo{author}{\bibfnamefont{R.~J.} \bibnamefont{Schoelkopf}},
  \bibnamefont{and} \bibinfo{author}{\bibfnamefont{P.}~\bibnamefont{Zoller}},
  \emph{\bibinfo{title}{Hybrid quantum processors: molecular ensembles as
  quantum memory for solid state circuits}} (\bibinfo{year}{2006}),
  \bibinfo{note}{submitted}.

\bibitem[{\citenamefont{Lukin}(2003)}]{lukin:457}
\bibinfo{author}{\bibfnamefont{M.~D.} \bibnamefont{Lukin}},
  \bibinfo{journal}{Reviews of Modern Physics} \textbf{\bibinfo{volume}{75}},
  \bibinfo{eid}{457} (pages~\bibinfo{numpages}{16}) (\bibinfo{year}{2003}).

\bibitem[{\citenamefont{Sachdev}(1999)}]{sachdev}
\bibinfo{author}{\bibfnamefont{S.}~\bibnamefont{Sachdev}},
  \emph{\bibinfo{title}{Quantum phase transitions}}
  (\bibinfo{publisher}{Cambridge University Press}, \bibinfo{address}{New
  York}, \bibinfo{year}{1999}).

\bibitem[{\citenamefont{Cirac et~al.}(1992)\citenamefont{Cirac, Blatt, Zoller,
  and Phillips}}]{cirac1992}
\bibinfo{author}{\bibfnamefont{J.~I.} \bibnamefont{Cirac}},
  \bibinfo{author}{\bibfnamefont{R.}~\bibnamefont{Blatt}},
  \bibinfo{author}{\bibfnamefont{P.}~\bibnamefont{Zoller}}, \bibnamefont{and}
  \bibinfo{author}{\bibfnamefont{W.~D.} \bibnamefont{Phillips}},
  \bibinfo{journal}{Physical Review A} \textbf{\bibinfo{volume}{46}},
  \bibinfo{pages}{2668} (\bibinfo{year}{1992}).

\bibitem[{\citenamefont{Lin et~al.}(2004)\citenamefont{Lin, Teper, Chin, and
  Vuletic}}]{lin:050404}
\bibinfo{author}{\bibfnamefont{Y.}~\bibnamefont{Lin}},
  \bibinfo{author}{\bibfnamefont{I.}~\bibnamefont{Teper}},
  \bibinfo{author}{\bibfnamefont{C.}~\bibnamefont{Chin}}, \bibnamefont{and}
  \bibinfo{author}{\bibfnamefont{V.}~\bibnamefont{Vuletic}},
  \bibinfo{journal}{Physical Review Letters} \textbf{\bibinfo{volume}{92}},
  \bibinfo{pages}{050404} (\bibinfo{year}{2004}).

\bibitem[{\citenamefont{Brown and Carrington}(2003)}]{browncarrington}
\bibinfo{author}{\bibfnamefont{J.~M.} \bibnamefont{Brown}} \bibnamefont{and}
  \bibinfo{author}{\bibfnamefont{A.}~\bibnamefont{Carrington}},
  \emph{\bibinfo{title}{Rotational Spectroscopy of Diatomic Molecules}}
  (\bibinfo{publisher}{Cambridge University Press}, \bibinfo{address}{New
  York}, \bibinfo{year}{2003}).

\bibitem[{\citenamefont{Childs et~al.}(1981)\citenamefont{Childs, Cok, Goodman,
  and Goodman}}]{childs}
\bibinfo{author}{\bibfnamefont{W.~J.} \bibnamefont{Childs}},
  \bibinfo{author}{\bibfnamefont{D.~R.} \bibnamefont{Cok}},
  \bibinfo{author}{\bibfnamefont{G.~L.} \bibnamefont{Goodman}},
  \bibnamefont{and} \bibinfo{author}{\bibfnamefont{L.~S.}
  \bibnamefont{Goodman}}, \bibinfo{journal}{Journal of Chemical Physics}
  \textbf{\bibinfo{volume}{75}}, \bibinfo{pages}{501} (\bibinfo{year}{1981}).

\end{thebibliography}

\end{document}